\documentclass[preprint,prb,showpacs,amsmath,amssymb]{revtex4}
\usepackage{graphicx}
\usepackage{dcolumn}
\usepackage{bm}

\begin{document}

\title{Absence of Pressure-Driven Supersolid Flow at Low Frequency}

\author{Ann Sophie C. Rittner}
\altaffiliation[current address: ]{Dept.\ of Physics and Astronomy and Rice Quantum Institute, Rice University, Houston, TX 77251}
\affiliation{Laboratory of Atomic and Solid State Physics, Cornell University, Ithaca, NY 14853-2501.}

\author{Wonsuk Choi}
\altaffiliation[permanent address: ]{Physics Dept., KAIST, Dejeon, South Korea} \affiliation{Laboratory of Atomic and Solid State Physics,
Cornell University, Ithaca, NY 14853-2501.}

\author{Erich J. Mueller}
\affiliation{Laboratory of Atomic and Solid State Physics, Cornell University, Ithaca, NY 14853-2501.}

\author{John D. Reppy}
\email[]{jdr13@cornell.edu}
\affiliation{Laboratory of Atomic and Solid State Physics, Cornell University, Ithaca, NY 14853-2501.}

\date{\today}

\begin{abstract}
An important unresolved question in supersolid research is the degree to which the non-classical rotational inertia (NCRI) phenomenon observed in the torsional oscillator experiments of Kim and Chan, is evidence for a Bose-condensed supersolid state with superfluid-like properties. In an open annular geometry, Kim and Chan found that a fraction of the solid moment of inertia is decoupled from the motion of the oscillator; however, when the annulus is blocked by a partition, the decoupled supersolid fraction is locked to the oscillator being accelerated by an AC pressure gradient generated by the moving partition. These observations are in accord with superfluid hydrodynamics. We apply a low frequency AC pressure gradient in order to search for a superfluid-like response in a supersolid sample. Our results are consistent with zero supersolid flow in response to the imposed low frequency pressure gradient. A statistical analysis of our data sets a bound, at the 68\% confidence level, of 9.6$\times 10^{-4}$~nm/s for the mass transport velocity carried by a possible supersolid flow. In terms of a simple model for the supersolid, an upper bound of 3.3$\times 10^{-6}$ is set for the supersolid fraction at 25~mK, at this same confidence level. These findings force the conclusion that the NCRI observed in the torsional oscillator experiments is not evidence for a frequency independent superfluid-like state. Supersolid behavior is a frequency-dependent phenomenon, clearly evident in the frequency range of the torsional oscillator experiments, but undetectably small at frequencies approaching zero.
\end{abstract}

\pacs{66.30.Ma, 67.80.bd}
\maketitle

Forty years have passed since the first suggestion, by G.V.Chester~\cite{Chester1970}, A.F. Andreev and I.M. Lifschitz~\cite{Andreev1969}, and A.J. Leggett ~\cite{Leggett1970}, that a Bose-condensed supersolid state might exist in solid $^4$He at sufficiently low temperatures. The remarkable discovery in 2004, by E. Kim and M.H.W. Chan (KC) \cite{Kim2004a,Kim2004}, of an anomalous drop in the rotational inertia of solid $^4$He in a torsional oscillator below 200~mK, provided the first convincing evidence for the existence of the supersolid state. The findings of KC have since been confirmed by other experimental groups~\cite{Kondo2006,Rittner2006,Aoki2007,Penzev2007}. In spite of extensive experimental work and vigorous activity by the theoretical community, the exact nature of the supersolid state remains unresolved at this time. A fundamental open question is whether the supersolid state is a Bose-condensed state with superfluid-like properties. In this letter, we address this question experimentally by examining the low frequency response of the supersolid to an imposed AC pressure gradient. 

KC performed an extensive series of experiments for both solid helium contained in porous media and for bulk samples contained in cylindrical and annular geometries. Annular geometry has the advantage of restricting the range of variation of the velocity field and by placing a blocking partition across the annulus also allows the topology of the sample to be altered from the doubly-connected sample geometry of an open annulus to a singly-connected geometry. In a blocked annulus experiment, Kim and Chan~\cite{Kim2004a} found that the magnitude of the supersolid response was reduced to 1~\% of the open annulus fraction. This ÒblockedÓ annulus result has also been confirmed by Rittner and Reppy~\cite{Rittner2008} and provides the strongest evidence, to date, that the supersolid phenomenon involves a long-range correlated flow similar to that of superfluid $^4$He.

When the annulus is blocked, the supersolid fraction is accelerated by a pressure gradient generated by the motion of the partition. This result suggests that it might be possible to observe a pressure-driven response in solid $^4$He in the temperature range of the supersolid state by means other than the torsional oscillator. In the years since the first suggestion of a supersolid state in solid $^4$He [1,2,3] there have been a number of unsuccessful attempts to observe pressure-driven supersolid flow~\cite{Bonfait1989,Greywall1977,Day2006}. The majority of these experiments search for evidence of supersolid mass-flow in response to a static pressure gradient. In a recent experiment, J. Day and J. Beamish~\cite{Day2006} were able to set a stringent limit on possible DC supersolid mass-flow. Their apparatus consists of two chambers filled with solid $^4$He separated by a microchannel plate. The narrow channels in the microchannel plate serve to lock the $^4$He solid and prevent, at low temperatures, plastic flow of the solid from one chamber to the other. When the volume of one chamber is reduced, an immediate pressure increase is seen in the second chamber. This pressure signal is attributed to a flexing of the microchannel plate separating the two chambers rather than to pressure-driven mass flow. Following the prompt pressure response, no further relaxation was observed for time intervals as long as 20 hours. This observation allows a stringent upper bound to be placed on any D.C. pressure-driven mass-flow from one chamber to the other. Presuming that any mass transport would be carried by flow of the supersolid fraction, the limit on the mass flow rate can be converted to a limiting flow velocity of the supersolid fraction. This limiting supersolid velocity will be inversely proportional to the supersolid fraction density. For a 1\% supersolid fraction, Day and Beamish obtain an upper bound of 1.2$\times 10^{-3}$ nm/s on any D.C. supersolid flow and a maximum displacement of 86 nm for such a flow over the 20~hour long observation period. The failure to find evidence for long-term DC supersolid mass transport could be expected because we know from the blocked annulus experiments that the supersolid can adjust to pressure gradients on millisecond time scales. Thus we expect the supersolid fraction to establish pressure equilibrium almost immediately as the static pressure gradient is initially imposed, and once this equilibrium is established, no further mass transport would be expected.

Would there be an observable response to a slow AC variation in the pressure gradient? If the supersolid does have superfluid-like properties, as might be expected for a Bose-condensed supersolid state, we would expect to see an AC response at low frequencies as well at torsional oscillator frequencies. In their experiment, Day and Beamish have searched unsucessfully for signs of AC pressure-driven supersolid response in their cell. Any AC supersolid response is superimposed on the feed-through signal arising from the flexing of the microchannel plate. Assuming a 1\% supersolid fraction and a limit on the superflow velocity of 2.1$\times 10^{-2}$ nm/s, Day and Beamish find an upper limit at 0.1~Hz of 3~nm for the maximum AC displacement.

\begin{figure}
\begin{centering}
\vspace{0.5cm}
\includegraphics[width=0.42\textwidth]{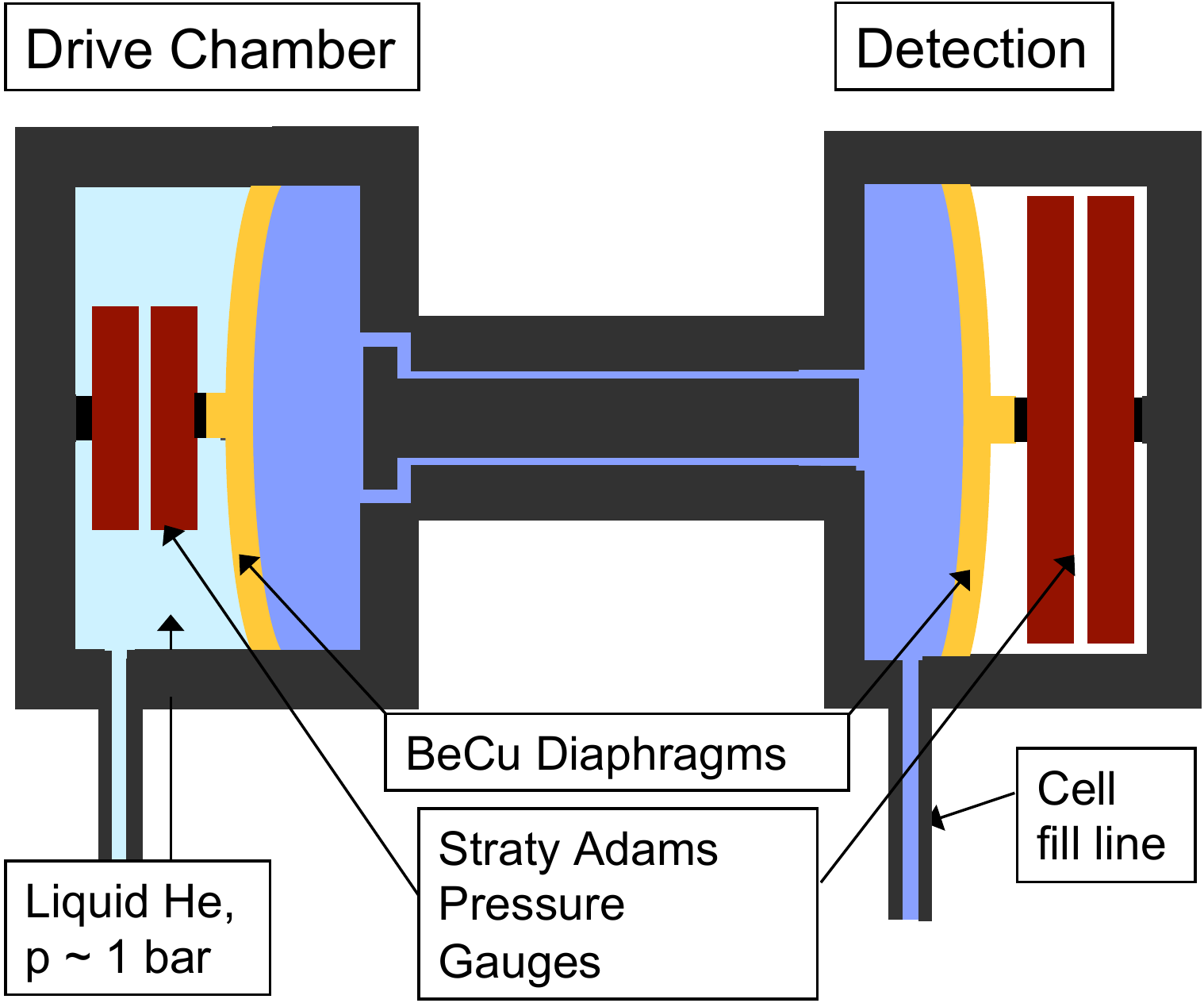}
\caption{\label{cartoon}Schematic diagram of BeCu flow cell. The total cell volume is 0.157~cm$^3$. Two Straty-Adams pressure gauges monitor the pressure of the drive and detection chamber. The pressure resolutions of the drive and detection capacitance gauges are 1~mbar and 2~$\mu$bar respectively. }
\end{centering}
\end{figure}

In an attempt to improve on these results, we have repeated the Day-Beamish experiment with a cell designed specifically to eliminate direct pressure feed-through, and in addition we have considerably increased the sensitivity of the detection system. A schematic view of our cell is shown in Figure~\ref{cartoon}. In our design, we have replaced the micro-channel plate of the Day and Beamish design with a narrow 4 cm long section consisting of a narrow annular slit to provide a connecting path of solid between two chambers. This narrow slit serves to lock the bulk solid in place and prevent plastic flow of the solid. The basic idea of the experiment is to squeeze the solid in one chamber and look for a supersolid mediated response in the second chamber. We have also increased the area of the detection capacitor in the interest of improved displacement sensitivity.

The internal spacing of the cell has been kept small to obtain a large surface-to-volume ratio, which has been shown in torsional oscillator experiments~\cite{Rittner2007} to be associated with large supersolid fractions. The solid samples are grown using the blocked capillary method to maximize the degree of disorder in the sample. The cell fill line consists of two meters of 0.1~mm internal diameter cupronickel tubing with heat sinks at a number of points between 4~K and the mixing chamber of our dilution refrigerator. The $^4$He used in these measurements is standard ultrapure grade commercial $^4$He with a nominal $^3$He concentration of 0.3~ppm. 

A narrow channel connecting the two chambers is created by clamping a 4~cm cylindrical rod, 1/8Ó in diameter, in a slightly oversized hole to form an annular slit with an average radial gap of 1.0$\times 10^{-2}$ cm and open area, $a_{chan}$ = 1.0$\times 10^{-2}$ cm$^2$. The end plates of the drive and detection chambers consist of flexible diaphragms, each with an area of a = 2.31 cm$^2$. The zero pressure height of the chambers has been machined to be close to 2.50$\times 10^{-2}$~cm. This height is increased by 0.13$\times 10^{-2}$~cm when the internal pressure is raised to 30~bar. At this pressure the volumes of the drive and detection chambers are estimated to be 0.0587~cm$^3$ and the total cell volume to be 0.157 cm$^3$. Parallel capacitor plates mounted on the center of each of the diaphragms serve to monitor the displacement of the end wall of each chamber. The separations between the plates for the drive and detection capacitors are calculated from the known areas of the capacitor plates and the capacitances measured by two A-D 2500A capacitance bridges. A correction is made for the dielectric constant of the liquid helium between the plates of the drive capacitor. The area of the detection capacitor plates is 5.07~cm$^2$, allowing a short-term pressure displacement resolution of about 2$\times 10^{-3}$~nm corresponding to a pressure change of a few $\mu$bar. The pressure sensitivity or spring constant of the detector diaphragm, $\kappa_2$ = $2.27 \times 10^4$~bar/cm, was determined by calibration at 4.2~K against an external pressure gauge. 

We apply an external force to the flexible diaphragm of the drive chamber via a controlled variable pressure source of liquid helium. This driving pressure is varied by heating a volume of liquid helium with an AC current supplied by a high resolution frequency synthesizer. This method has the advantage of conveniently producing pressure swings of the desired amplitude on the order of fractions of a bar, but unfortunately it was restricted in our current design to frequencies below 0.1~Hz. The variable pressure generated in the heated volume is transmitted to the experimental cell through a liquid-filled fill line similar in construction to that used to fill the cell. The applied liquid pressure is monitored with a Straty-Adams pressure gauge~\cite{Straty1969} at the mixing chamber (not shown in Figure \ref{cartoon}).

\begin{figure}
\begin{centering}
\includegraphics[width=0.51\textwidth]{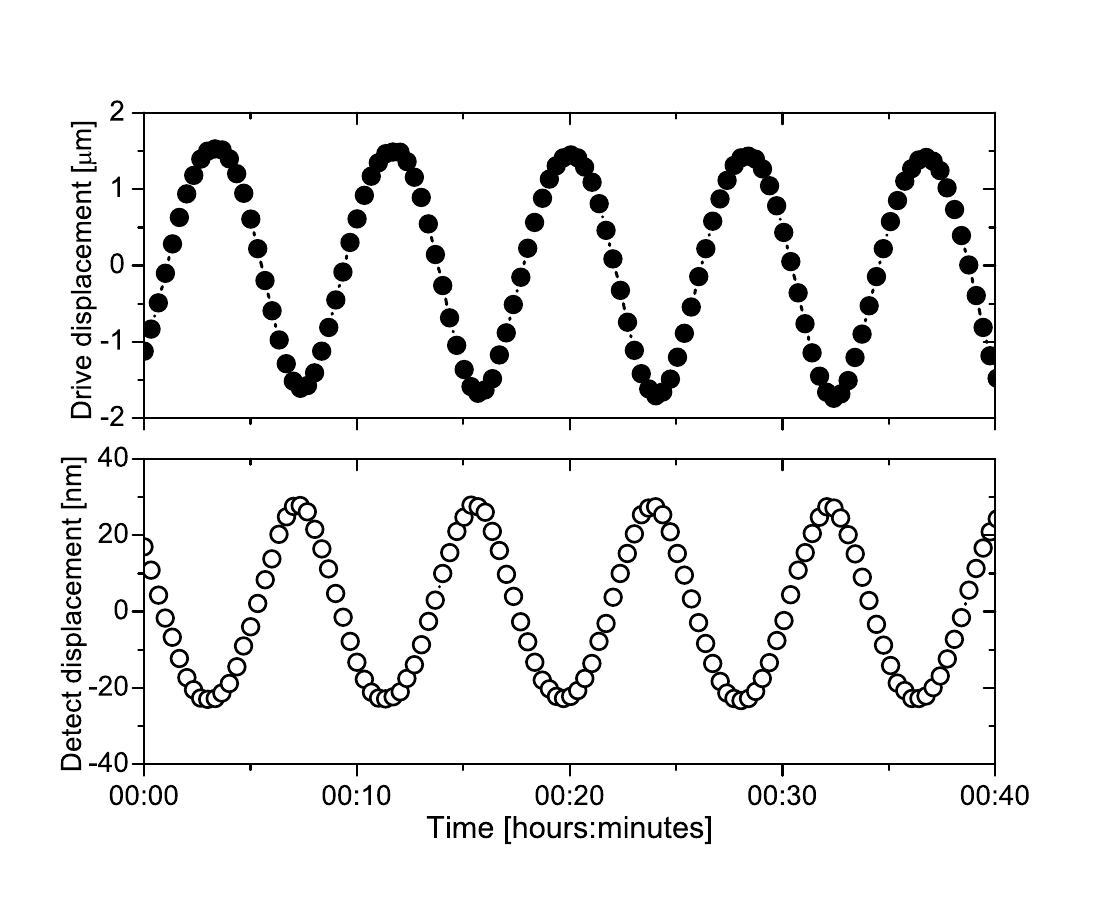}
\caption{\label{liquidsignal} The displacement of the drive and detection diaphragms are shown as a function of time for a cell filled with liquid $^4$He. The drive and detection displacements have a phase difference of 180$^{\circ}$ because the drive pressure is applied to the diaphragm of the drive chamber. The liquid pressure is 30.3~bar, the temperature is stabilized at 1.90~K and the drive frequency is 2~mHz.}
\end{centering}
\end{figure}

Figure \ref{liquidsignal} shows displacement of the drive and detection diaphragms in response to a slow 2~mHz AC variation in the external drive pressure while the cell was filled with normal phase liquid helium. At a pressure of 30.3~bar and a temperature of 1.90~K the liquid sample is at a point in the pressure-temperature phase diagram close to the melting curve. The liquid sample is sealed off by holding a section of the fill line at a temperature below the melting curve. The displacements of the diaphragms are monitored as changes in the separation of the capacitor plates. Our convention has been to count an increase in capacitor plate separation as a positive diaphragm displacement. Thus, with an increase in the applied drive pressure, the drive chamber diaphragm will be displaced to the right (see Fig/. \ref{cartoon}) and the capacitor plate separation will increase. The reduction in the total cell volume produced by this motion of the drive diaphragm leads to an increase in the overall cell pressure and a reduction in the capacitor spacing on the detection side. If we assume that both diaphragms are rigidly clamped at their edges, then the total volume, $\Delta V$, displaced by the motion the diaphragms is $\Delta V = - (a/3)( d_1 + d_2)$, where a is the area of each diaphragm and $d_1$ and $d_2$ are the respective displacements of the drive and detector diaphragms. The minus sign is a consequence of our convention in designating diaphragm displacements. The ratio of the displacements is $(d_1/d_2 ) = - [1 + (3V \kappa_2 K/a)]$, where K is the compressibility of the liquid and V is the total volume of the liquid sample contained in the cell plus the fill line and any of the sintered copper heat sinks between the cell and the point where the fill line is frozen. In the case of an incompressible fluid, K = 0, and the displacement of the diaphragms will be equal in magnitude but opposite in sign. Liquid and solid helium, however, are both rather compressible and the term $(3V \kappa_2 K/a) >> 1$. For the data plotted in Figure \ref{liquidsignal}, the amplitude of displacement for the drive diaphragm is $d_1$ = 1.6$\times 10^{-4}$~cm while the detection diaphragm has a displacement, $d_2$ = - 2.46$\times 10^{-6}$ cm for a liquid sample pressure amplitude of 55.8~mbar. The ratio $(d_1/d_2)$ = - 67 is in reasonable agreement with a value calculated based on the estimated of the volume of the low temperature liquid sample, a value for the liquid compressibility of 3.45$\times 10^{-3}$ bar$^{-1}$\cite{Edwards1965}, and measured value of the spring constant $\kappa_2$. The maximum volume displaced by the motion of the drive diaphragm is $\Delta V_1 = (a d_1)/3$ = 1.23$\times 10^{-4}$ cm$^3$. If this entire volume were to flow back and forth through the slit between the two chambers, the maximum flow velocity would be $v_s = 2\pi f (\Delta V/a_{chan}) = 1.55$~$\mu$m/s, a value below the typical torsional oscillator critical velocity of 10~$\mu m$/s. 

Although the data shown in Figure \ref{liquidsignal} were taken in the normal phase liquid helium, there would be little difference if the cell were filled with superfluid. The viscosity of the normal fluid is small, on the order of 200 micropoise for the given temperature and pressure, and consequently the maximum pressure drop for Poisuelle flow through the slit is negligible compared to the pressure amplitude, $\Delta P = 55.8$~mbar arising from the compression of the liquid sample.

\begin{figure}
\vspace{-.2cm}
\begin{centering}
\includegraphics[width=0.51\textwidth]{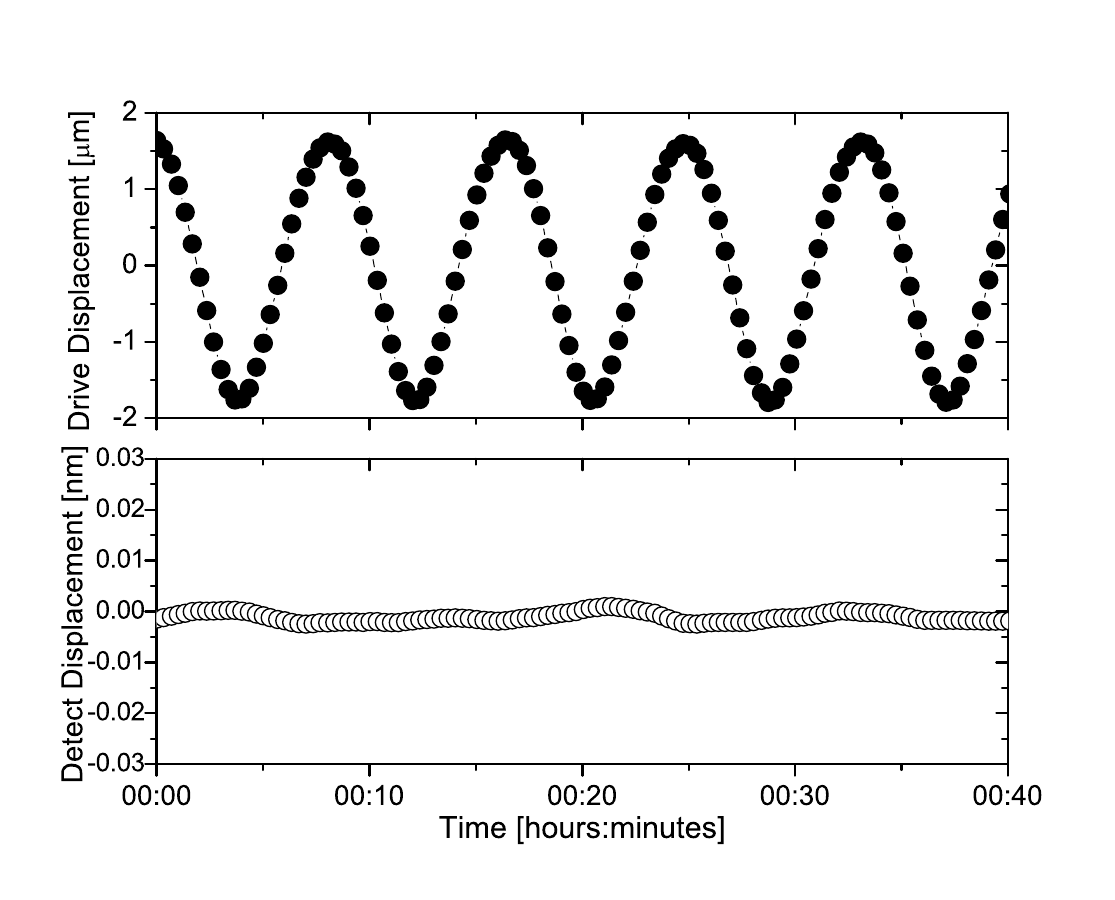}
\caption{\label{solidsignal} The displacements of drive and detection diaphragms are shown as a function of time when the cell is filled with solid $^4$He at a pressure of 30~bar. The cell temperature is stabilized at 20~mK and the drive frequency is 2~mHz. }
\end{centering}
\end{figure}

When the low temperature system is filled with solid helium, the relevant volume is now the cell volume alone, since we do not expect any appreciable mass flow of solid through the 0.1 mm ID fill line. In Figure 3 we plot a typical set of drive and detection diaphragm displacement data with the cell filled with solid. Since the supersolid signal reaches its maximum value at the lowest temperatures, we would expect to see the largest response in our experiment at the lowest temperatures. The data set shown in this figure were taken over a relatively short period, 40 minutes, at a pressure of 30~bar and controlled at the low temperature of 20~mK. Assuming that little mass leaves the drive chamber, the displacement amplitude of the drive diaphragm, $d_1$ = 1.75$\times 10^{-4}$~cm, corresponds to a pressure swing of 0.74~bar. If mass transport were to take place in response to the periodic variation in the drive pressure, we would expect to see a correlated displacement, $d_2$, of the detection chamber diaphragm as mass flows in and out of the detection chamber. The striking feature of these data, however, in contrast to the situation when the cell was filled with liquid, is the absence of any signal, discernable to the eye, above a 2$\times 10^{-3}$~nm noise level. We have repeated this type of measurement for a number of solid samples formed at different pressures ranging from 27 to 40~bar and for fixed temperatures ranging from 20 to 400~mK. In no case is there any discernable indication, above noise, of a signal indicative of AC pressure-driven mass-flow between the two chambers of our cell. Therefore, we conclude that within the resolution of the data shown in Figure 3, that the product, $(\rho_s/\rho)v_s$ is consistent with zero, where $(\rho_s/\rho)$ is the supersolid fraction and $v_s$ is the flow velocity of the supersolid fraction in the narrow channel connecting the two chambers. 

\begin{figure}
\vspace{.7cm}
\hspace{-.4cm}
\begin{centering}
\includegraphics[width=0.47\textwidth]{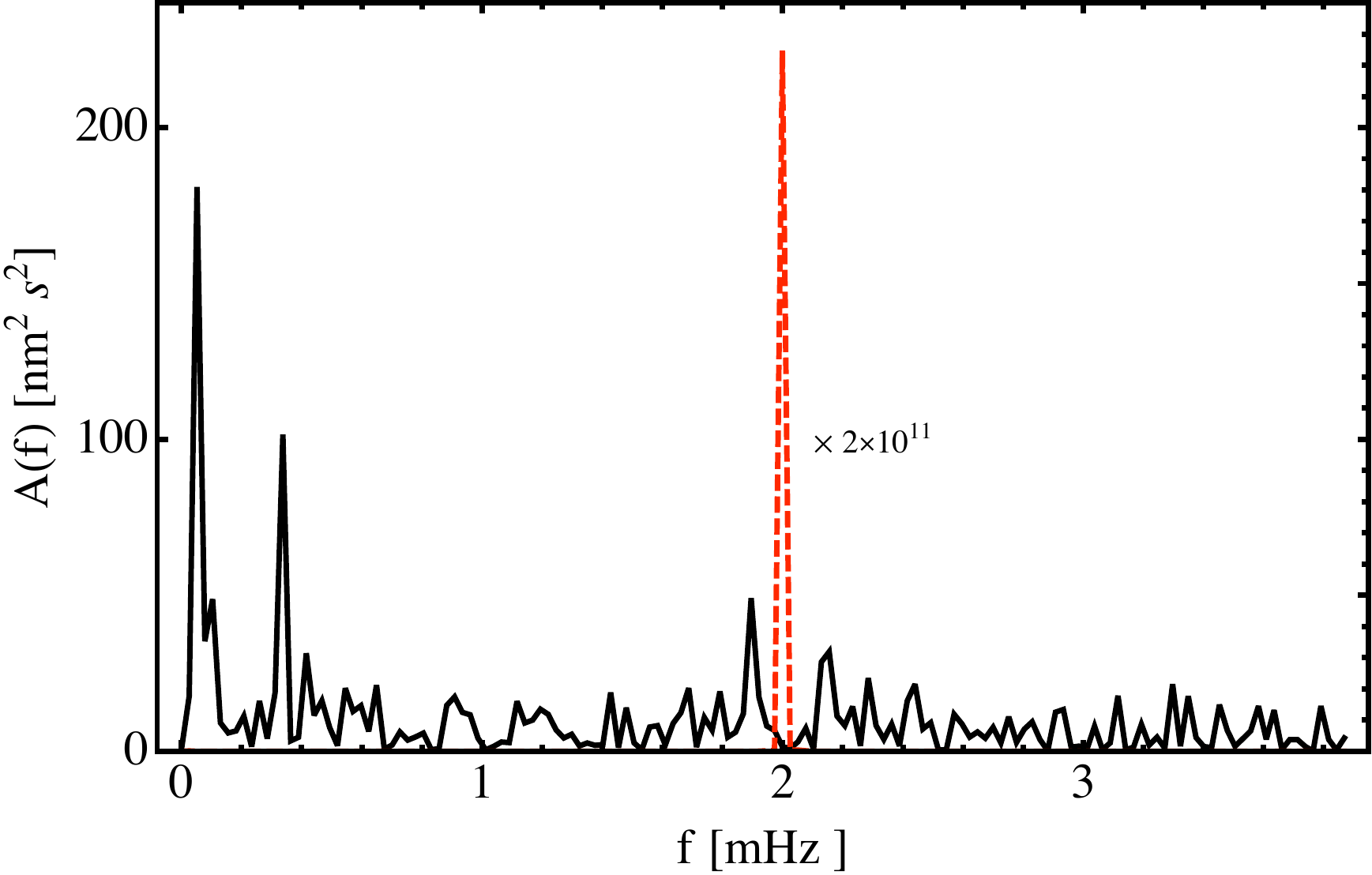}
\caption{\label{4spectraldensity} Spectral density of drive (dashed line) and detection amplitude (solid line) for 77 cycles. The cell pressure is 30.3~bar and the temperature is stabilized at 25~mK.}
\end{centering}
\end{figure}

We now proceed to a more quantitative analysis of our data. As an example, we shall present an analysis of a set of a 10~hour time period (2000 data points) where the cell temperature was held fixed at 25$\pm 0.2$~mK and at a pressure of 30.3~bar. As a first step, we compute the power spectra for both the drive and the detection signals. The spectra are shown in Figure \ref{4spectraldensity}. The spectrum for the drive (dashed line) shows the expected sharp peak at 2~mHz, the spectrum for the detection signal (solid line) shows noise with no evidence of a signal at 2~mHz.

In a further analysis of this data set, we find that the mean displacement amplitude of the drive diaphragm is $d_1 $= 0.365~$\mu$m with a drive pressure amplitude of 0.153~bar. To produce a sharp statistical bound on the response at 2~mHz, we divide the data into 19~segments, each of which is 4~cycles long and contain 107~data points. By looking at the 2~mHz Fourier component of the jÕth subinterval, we calculate the best fit sinusoidal curve, $d^{(j)}(t)=(A_je^{i\omega t}+A_j*e^{-i\omega t})/2$, where $\omega/2\pi$ = 2~mHz. The amplitude and phase of the response during that segment is $|A_j|$ and $arg(A_j)$. We choose our origin of time so that the drive has a phase of 0, and we expect the response to have a phase of 180¡. Figure 4 shows a scatter plot of the coefficients of the response, with an inset showing the drive. We find that the mean amplitude of the response is $|A| = 8.0\times10^{-5}$~nm and the mean phase is $\phi$ = 75$^{\circ}$. The standard deviation of these 19 segments is $\sigma = 7.8\times10^{-4}$~nm, yielding a standard error of the mean of $\sigma_e$ = $1.8\times10^{-4}$~nm. The latter is the statistical estimate of how much the calculated mean from these 19 segments is expected to differ from the mean of an infinite number of cycles. Statistically, our result is consistent with zero response. More precisely, we bound the amplitude to be less than $|A|+\sigma_e$ = $2.6\times10^{-4}$~nm, with roughly a 60\% confidence. The value of $\sigma_e$ = 1.8$\times 10^{-4}$ nm, obtained for this fit, would correspond to pressure change, within the detection chamber of $\Delta P_2$ = 4.1$\times 10^{-7}$ bar, which could arise from the flow an additional mass, $m = \rho(K \Delta P_2 V_2)$, into the volume $V_2$ of the detection chamber. The maximum flow velocity in the channel, $v = \omega[(K \Delta P_2 V_2)/a_chan]$ = $(\rho_s/ \rho)v_s$. At the frequency $\omega$ and for a solid compressibility at 30 bar of 3.1$\times 10^{-3}$ bar$^{-1}$, v = 9.6$\times 10^{-4}$ nm/s. Since $v = (\rho_s/ \rho)v_s$, one must specifiy a supersolid fraction to obtain the supersolid flow velocity; for instance, for a supersolid fraction of 1\%, the one $\sigma_e$ bound would correspond a superflow velocity of 9.6$\times 10^{-2}$ nm/s. If the supersolid fraction, however, were to flow through the channel at 10 $\mu$m/s, the typical critical velocity seen in the torsional oscilltor experiments, then the supersolid fraction would be no larger than 2.4$\times 10^{-8}$, a value four orders smaller than the minimum value, $(\rho_s/\rho) $ = 3.0$\times 10^{-4}$, reported for torsional oscillator experiments~\cite{Clark2007}.

With additional model assumptions, one can obtain an estimate for the supersolid fraction corresponding to a given displacement of the detection diaphragm without a direct reference to the flow velocity. In this analysis we shall assume the elastic modulus, $M = K^{-1}$, of the solid can be considered as the sum of two parts, a contribution, $(\rho_s/\rho) K^{-1}$, corresponding to the supersolid fraction and a contribution, $ [1 - (\rho_s/\rho)] K^{-1}$, from the remaining solid. We shall also assume that the partial pressure of the supersolid fraction can achieve equilibrium independent of pressure gradients that might exist in the non-supersolid fraction. These are the conditions that would hold for a pure superfluid contained in a compressible porous medium. Under these assumptions, we can obtain an expression for the ratio of the displacements of the two diaphragms, $(d_1/d_2) = - [1 + (\rho/\rho_s) Ð 1) (V/V_2) + (\rho/\rho_s) (3V\kappa2 K/a)]$. This expression is then a generalization of the earlier relation for the single-phase fluid. Solving for the supersolid fraction we have $(\rho_s/\rho) = [(V/V_2) + (3V\kappa2 K/a)]/[(V/V_2) Ð (1 + (d_1/d_2))]$. If we set $(d_1/d_2) = -(d_1/ \sigma \epsilon ) = 2.0 \times 10^5$ then $(\rho_s/\rho)$ = 3.3$\times 10^{-6}$. This value for the supersolid fraction is nearly two orders of magnitude smaller than the minimum value reported for torsional oscillator experiments~\cite{Clark2007}.

In an earlier experiment, Y. Aoki, J.C. Graves, and H. Kojima [8] explored the frequency dependence of the supersolid response for an open cylindrical sample in a two-frequency (496 and 1173 Hz) torsional oscillator. Although they found a small reduction in the supersolid signal for the low frequency mode at temperatures above 35 mK, they found no difference for temperatures below this value. In contrast, in more recent experiments with an annular two-frequency cell, the Kojima group~\cite{Keiderling2009} found a 10\% reduction for the low frequency mode that persists to their lowest temperature. This may be an important finding and points to the need to extend torsional oscillator measurements to even lower frequencies. An important challenge for the future will be to devise experiments that span the $10^5$ range between the lowest torsional oscillator experiments (185 Hz)~\cite{Rittner2006} and the mHz measurements reported here.

In contrast to the absence of evidence for pressure-driven supersolid flow a zero or low frequencies in the experiments discussed thus far, there are two positive reports of mass transport between two regions occupied by liquid through an intervening region containing solid $^4$He. The first of these is an experiment on the melting curve by S. Sasaki, F. Caupin, and S. Balibar~\cite{Sasaki2007} at LÕEcole Normal Superiere Paris. Gravitationally driven flow was occasionally observed between two different regions of liquid helium separated by an intervening region of solid. In the second of these experiments, reported by M.W. Ray and R.B. Hallock~\cite{Ray2008} at the University of Massachusetts, mass transport was observed through a solid region at pressures a few bar above the melting curve. In their experiment, Ray and Hallock take advantage of the fact that helium can remain in the fluid phase within the pores of Vycor glass at pressures well above the melting curve. Fluid is introduced and extracted from the sample cell through two Vycor rods. The mass transport observed in both of these experiments is relatively pressure independent as would be characteristic of superflow limited by a critical velocity. A second feature of these experiments is the observation of the mass-flow at temperatures well above the temperature region of supersolid behavior as observed in torsional oscillator measurements.

How then can we reconcile our findings with the positive results of the Paris and Massachusetts groups? There has a suggestion by Caupin, Sasaki and Balibar~\cite{Caupin2007}, that the mass transport observed on or near the melting curve may, in fact, be the flow of ordinary superfluid $^4$He through a connected network of channels formed at the grain boundaries of polycrystalline samples or at the wall of the cell. Flow along grain boundaries or dislocation lines might account for the mass transport in the Massachsetts experiments. An alternate nonsuperfluid explanation might be based on the Òfrost heaveÓ phenomenon. Growth of the solid can occur at the fluid-solid interface at the surface of the vycor rod when the pressure of the superfluid in this rod is raised above the equilbrium value for the bulk solid. Extrusion of this newly formed solid into the bulk region would produce pressure gradients, elastic displacement, and possibly plastic flow of the bulk resulting in mass transport. The rate of such growth would be expected to be limited by thermal conduction as the heat of fusion enters and leaves the liquid solid interface and would be nearly constant in time, thus the frost heave mechanism could mimic the constant mass-flow expected for critical velocity superflow. If these proposed mechanisms were to provide correct explanations for the Paris and UMass observations, then the mass flow observed in these two experiments is not relevant to the supersolid phenomenon. 

\begin{figure}
\vspace{.7cm}
\hspace{-.4cm}
\begin{centering}
\includegraphics[width=0.47\textwidth]{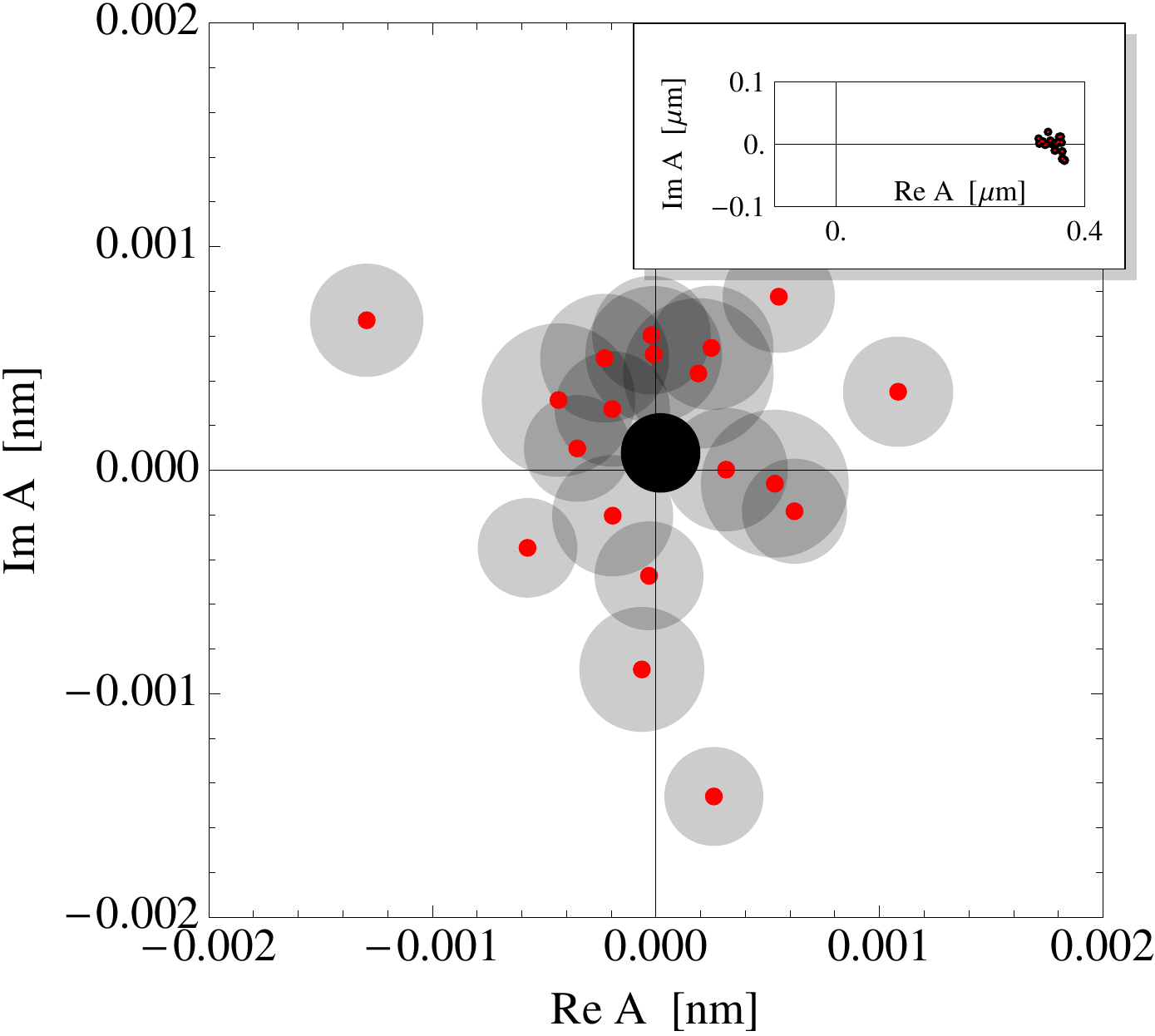}
\caption{\label{spectraldensity} Amplitude of detector response. Small (red) dots represent the best fit amplitude and phase of the motion of the diaphram in the detection chamber at $\omega/2\pi=2$~mHz [$d_2=\Re[A \exp^{i \omega t}] ]$ over a four cycle period. The diffuse gray circle around each dot represents the statistical estimate of the standard error in A. The large black dot near the origin represents the average of the 19 data points. The size of the black dot corresponds to the standard error in the mean, and represents the region over which one has a 60\% confidence that after an infinite number of cycles the mean would be in that region. The inset on the upper right depicts the same quantity for the drive.}
\end{centering}
\end{figure}

We briefly discuss the various theories of supersolidity, and how our observations dramatically reduce the number of viable scenarios. The first class of theories are based upon Bose-Einstein condensation of vacencies. The simplest such theories \cite{Andreev1969,Chester1970} consider a dilute gas of vacencies in an otherwise perfect crystal, while more sophisticated models have these vacencies living in extended defects \cite{Biroli2008,Boninsegni2007,Pollet2007,Balatsky2006,Khairallah2005}. A large amount of numerical \cite{Pollet2008,Prokofev2007,Boninsegni2006,Boninsegni2006a,Clark2006,Galli2005,Ceperley2004,Vitiello1988} and theoretical \cite{Saslow2009,Anderson2009,Dorsey2006,Baskaran2009,Tiwari2004,Anderson2005, Dai2005} work, has been devoted to exploring these models, both from phenomenological and microscopic perspectives. On a macroscopic scale, most of these models are equivalent to an effective description in terms of a superfluid in a porous medium. Our observations, and analysis, effectively rule out any such description.

The second class of theories are purely mechanical in nature. For example Yoo and Dorsey analyzed the possibility that a viscoelastic model could account for the torsional oscillator measurements, finding that they could account for the dissipation in the torsional oscillators, but not for the frequency shift \cite{Yoo2009}. Both Andreev \cite{Andreev2007} and Balatsky et al. \cite{Balatsky2007} argued for glassy models. Similarly, Nussinov et al. suggested that the freezing in of glassy regions could be responsible for most of the observations \cite{Nussinov2007}, but further experimental studies have cast doubt on that explanation \cite{Clark2006}. These mechanical models typically lead to strong frequency dependancies, and can readily be made consistent with our observations. Regardless of their virtues, it is hard to reconcile a non-superfluid scenario with the blocked annulus experiments, but they are undoubtedly responsible for similar observations in solid hydrogen \cite{Clark2006}

Recently Hunt et al. \cite{Hunt2009} proposed a hybrid scenario where superfluidity is controlled by the dynamics of some glassy degrees of freedom. While largely phenomonological, this model accomodates both strong frequency dependance (from the glassy degrees of freedom) and the ability to support supercurrents, and may be consistent with our observations. In a similar vein, Anderson has proposed that supersolidity might be due to a vortex liquid state \cite{Anderson2007}.
Shimizu et al. \cite{Shimizu2009} have interpreted their torsional oscillator measurements in this manner, and Chan \cite{Chanreview2008} argues that the measurements of Aoki et al. \cite{Aoki2007} could also be explained by this theory. A vortex liquid would have very pronounced frequency dependance in its response, with vanishing superfluid response at zero frequency. This would be consistent with our observations.

Based on the null result of our AC pressure-driven flow measurements and the earlier results of the Alberta group, we conclude that the supersolid phenomenon does not obey the conventional hydrodynamics of a superfluid, but rather exhibits frequency-dependent behavior, with strong supersolid signals observable at typical torsional oscillator frequencies between a few hundred and a few thousands of Hz, but is absent or unobservable at zero or low frequencies. 

\begin{acknowledgments}
The authors would like to thank V. Ambegaokar, N.W. Ashcroft, D.M. Lee, M.H.W. Chan and E. Kim for useful and encouraging advice. This work has been supported by the NSF DMR-060584, PHY-0758104 and by CCMR through NSF DMR-0520404. In addition W. Choi wishes to acknowledge financial support through Creative Research Initiatives (Center for Supersolid \& Quantum Matter Research) of MEST/KOSEF.
\end{acknowledgments}

\end{document}